\documentclass[pra,amssymb,amsfonts,showpacks,twocolumn,showpacs]{revtex4}
\usepackage{amsmath}
\usepackage{amsfonts}
\usepackage{amssymb}
\usepackage{bm}

\def\beq{\begin{equation}}
\def\eeq{\end{equation}}
\def\bea{\begin{eqnarray}}
\def\eea{\end{eqnarray}}

\def\half{\mbox{$1\over2$}}

\def\pad{\partial}
\def\tr{{\rm tr}}

\def\rc{{\rm c}}
\def\rq{{\rm q}}

\def\cH{{\cal H}}

\def\1{\mbox{1\hskip-.25em l}}
\def\6{\langle }
\def\9{\rangle}

\begin{document}

\title{Inconsistency of  quantum--classical dynamics, and what it implies}

\author{ Daniel R. Terno}
\affiliation{Perimeter Institute for Theoretical Physics, 35 King St. N., Waterloo,
Ontario, Canada N2J 2W9}

\begin{abstract}
A new proof of the impossibility of a universal quantum-classical
dynamics is given.  It has at least two consequences. The standard
paradigm ``quantum system is measured by a classical apparatus" is
untenable, while a quantum matter can be consistently coupled only
with a quantum gravity.
\end{abstract}
\pacs{03.65.Sq, 03.65.Ta, 04.60-m}
\maketitle
%\section{Introduction}
Quantum mechanics gives exceedingly accurate predictions for
atomic and nuclear systems. Classical mechanics is just as
successful for planetary motion. When it is necessary or
convenient to  describe part of the system classically and another
part quantum mechanically both languages are combined. The
phenomena that are treated by this mixed dynamics range form the
one-loop quantum gravity \cite{gra} and quantum cosmology
\cite{cosm} to gas kinetics
\cite{gas} and dynamics of chemical reactions \cite{chem}.
Different mixed dynamics schemes are adopted for different
situations. Often they produce a good agreement with experimental
data or reasonable physical models, but these schemes are also
prone to sudden breakdowns and require fine tuning of parameters
or a careful selection of allowed initial states.

Quantum mechanics itself is dependent on such a coupling, because
its basic notions are ``preparations'' and ``tests''
(measurements). They are performed by macroscopic devices and
these are described in classical terms. Any intermediate systems
used in that process could be treated quantum mechanically, but
the final instrument has a purely classical description. The
necessity of using a classical terminology was emphasized by Bohr
\cite{bohr}, even though he never considered the measuring process
as  dynamical interaction between an apparatus and the system
under observation. Yet, measuring apparatuses are made of the same
kind of matter as everything else and they obey the same physical
laws. It therefore seems natural to use quantum theory in order to
investigate their behavior during a measurement. This was
attempted by von Neumann
\cite{vn}. Von Neumann represented the apparatus by a single degree of
freedom, whose value was correlated to that of the dynamical
variable being measured. Such an apparatus is not, in general,
left in a definite pure state and does not admit a classical
description. Therefore, von Neumann introduced a second apparatus,
which observes the first one, and possibly a third apparatus, and
so on, until there is a final measurement, which is not described
by quantum dynamics and has a definite result. These different
approaches of Bohr and von Neumann are reconciled by a dual
description of the measuring apparatus. It obeys quantum mechanics
while it interacts with the  system under observation, and then it
is ``dequantized'' and is described by a classical Liouville
density which provides the probability distribution for the
results of the measurement \cite{hp,diosi}.

Is it possible to consistently maintain a distinction between a
classical apparatus and a quantum system during {\em all} stages
of the measurement process? Is it possible to consistently argue
about the reaction of a quantized matter on a classical metric?
Both answers depend on the existence of a general
quantum-classical interaction scheme, as opposed to the different
effective methods that were mentioned above. A universal mixed
dynamics formalism should be similar to, e.g., the Hamiltonian
formulation of classical and quantum mechanics, where the
information about a system is provided by its Hamiltonian, which
is then plugged into the standard scheme. Such universal formalism
is impossible
\cite{casa,pt}. In this paper we give a new derivation of this
result from a minimal number of assumptions and explore its
consequences.  Koopmanian formalism
\cite{koopman,qt,mau} for classical mechanics  illustrates both
the desiderata of the measurement description and helps to derive
the impossibility of the mixed dynamics in a concise form.

%\section{Koopmanian formalism}

 For
simplicity we consider a single degree of freedom and denote the
canonical variables as $x$ and $k$  (we  reserve the symbol $p$
for the momentum of a quantum system, to be introduced later). Let
us write the Liouville equation as
\beq
i\,\partial f/\partial t=Lf, \label{Leq}
\eeq
where $L$ is the Liouville operator, or Liouvillian,
\beq
L=\left({\partial H\over\partial k}\right)
 \left(-i{\partial\over\partial x}\right)
 -\left({\partial H\over\partial x}\right)
 \left(-i{\partial\over\partial k}\right). \label{L}
\eeq
The Liouville density $f$ is never negative. It is convenient to
introduce likewise a ``classical wave function''
\beq
\psi_\rc\equiv\sqrt{f},\label{claswave}
\eeq
 which in this case satisfies the same equation of motion as $f$,
\beq
i\,\partial\psi_\rc/\partial t=L\psi_\rc.
\eeq
We shall now consider $\psi_\rc$ as the fundamental entity (but
only $f=|\psi_\rc|^2$ has a direct physical meaning). It can be
proved that, under reasonable assumptions about the Hamiltonian,
the Liouvillian is an essentially self-adjoint operator and
generates a unitary evolution~\cite{reed}:
\beq
\6\psi_\rc|\phi_\rc\9:=\int\psi_\rc(x,k,t)^*\,\phi_\rc(x,k,t)\,dxdy={\rm const.}
\eeq
It is possible to further mimic quantum theory by introducing {\it
commuting\/} operators $\hat{x}$ and $\hat{k}$, defined by
\beq
\hat{x}\,\psi_\rc=x\,\psi_\rc(x,k,t)\qquad{\rm and}
  \qquad\hat{k}\,\psi_\rc=k\,\psi_\rc(x,k,t).
\eeq
Note that the momentum $\hat{k}$ is not the shift operator (the
latter is $\hat{p}_x=-i\pad/\pad x$).  Likewise the boost operator
is $\hat{p}_k=-i\pad/\pad k$. These two operators are not
observable. We shall henceforth omit the hats over the classical
operators when there is no danger of confusion.

The analogy with quantum mechanics can be pushed further. What we
have above is a ``Schr\"odinger picture'' (operators are constant,
wave functions evolve in time as $\psi(t)=U(t)\psi(0)$, where
$U(t)=e^{-iLt}$ if the Hamiltonian is time-independent). We can
also define a ``Heisenberg picture'' where wave functions are
fixed and operators evolve:
\beq
X_H(t)=U^\dagger XU.
\eeq
The Heisenberg equation of motion,
\beq
i\,dX_H/dt=[X_H,L_H]=U^\dagger[X,L]\,U,
\eeq
together with the Liouvillian (\ref{L}), readily provide
Hamilton's equations
\beq
{dx\over dt}={\pad H\over\pad k},\qquad\qquad
     {dk\over dt}=-{\partial H\over\partial x}.
\eeq
%There is however an important difference: the time translation
%operator $L$ is not the energy, and its spectrum may extend to
%$-\infty$ \cite{qt,pt}.

This formalism allows to describe the states of classical  and
quantum systems in a single mathematical framework, namely in the
joint Hilbert space $\cH=\cH_\rq\otimes\cH_\rc$.
 Since we are dealing
with the Hilbert spaces, the concepts of a partial trace and
entanglement (including the one between classical and quantum
states) are naturally defined.

%\section{``Derivation" of the measurement formalism}
 A measurement
apparatus produces a classical (i.e., robust, clonable) data. If
the measurement interaction is sufficiently well understood, the
resulting quantum state is given by a set of Kraus matrices
\cite{cond}. Consider any measurement that can result in a finite
number of the discrete outcomes (this is what happens in any
actual experiment). If a particular outcome $\mu$ is registered,
then
\beq
\rho_{|\mu}=\frac{1}{\tr(\rho E_\mu)}\sum_i A_{\mu i}\rho A_{\mu
i}^\dag,\label{tran}
\eeq
where $\rho$ is the initial density matrix, a probability of the
outcome $\mu$ is calculated as an expectation of a positive
operator-valued measure (POVM) element $E_\mu$,
\beq
p(\mu|\rho)=\tr\rho E_\mu,
\eeq
and the POVM is constructed from the Kraus matrices $A_{\mu i}$,
\beq
E_\mu=\sum_i A_{\mu i}^\dag A_{\mu i}\label{povm}.
\eeq

To literally accept the paradigm ``quantum system-classical
apparatus", we need a description of dynamics that leads to the
above probability formula and conditional evolution. It is well
understood that the decoherence \cite{zurek} plays an important
role in the measurement process. Hence, in addition to the
Hamiltonian interaction between system and apparatus an
interaction with the environment should be taken into account.
This interaction typically selects a preferred {\em pointer basis}
and severely limits possible POVMs \cite{zurek}.

To mimic a quantum formalism, we describe classical measurements
by projection operators $\Pi_\mu$. Since any Liouville density can
be put into the form of a pure state by Eq.~(\ref{claswave}),
\beq
\Pi_\mu\equiv|\psi_{\rc \mu}\9\6\psi_{\rc \mu}|,
\eeq
where $\6\psi_{\rc\mu}|\psi_{\rc \nu}\9=0$, $\mu\neq\nu$. This set
may be taken as  complete,
\beq
\sum_\mu \Pi_\mu=\1.
\eeq
For example, we can partition the phase space into  domains
$X_\mu$. Then the projectors may be taken as
\beq
\Pi_\mu(x,k)=\chi_\mu(x,k), \label{pmes}
\eeq
where $\chi_\mu$ is an indicator of the set $X_\mu$ (equals to 1
if $(x,k)\in X_\mu$ and is zero otherwise). Hence
\beq
\tr(\Pi_\mu|\phi_\rc\9\6\phi_\rc|)=\int_{X_\mu}\!|\phi(x,k)|^2dxdk
\eeq
It should be noted that projectors onto some orthogonal basis
states of the classical Hilbert space do not necessary have a
physical meaning. Similarly, while the state on, say, a two
particle Hilbert space $\cH\otimes\cH'$ may be written in the form
$\psi_{12}=\sum_i\phi_\rc^i(x,k){\phi_\rc^i}'(x',k')$, the
apparent entanglement is fictitious. It can't be detected by any
measurement of the type of Eq.~(\ref{pmes}).

%\subsection{A wish list for the measurement}
Now we can state how the measurement formalism should be derived.
We follow the standard splitting of the Universe into a triple:
the system -- the apparatus -- their environment
\cite{qt,p:01,zurek}. The system is described by quantum mechanics,
the apparatus by classical mechanics, and the environment may be
anything. For classical systems we use the Koopmanian formalism.
An initial state is
\beq
\rho_\rq\otimes\rho_\rc\otimes\alpha,
\eeq
where $\alpha$ is an initial state of the environment. The overall
evolution is unitary. It may entangle anything with anything else,
but it is reversible. The irreversibility is introduced by tracing
out the environment, which may also dictate the choice of the
basis.
 An outcome $\mu$ of the measurement induces the state transformation
\beq
\rho_{\rq|\mu}=\frac{1}{p(\mu|\rho)}(\1_\rq\otimes\Pi_\mu)\tr_{\rm env't}\left[
U(\rho_\rq\otimes\rho_\rc\otimes\alpha)U^\dag\right]
(\1_\rq\otimes\Pi_\mu),
\eeq
and by comparison with Eqs.~(\ref{tran})-(\ref{povm}) the POVM and
Kraus matrices can be identified.

If we ignore the environment, the (pre-)measurement
\cite{qt} process may be described as follows.  Initially the
combined system is in the state
\beq
|\Psi^0\9=|\psi_\rq^0\9\otimes|\psi_\rc^0\9,
\eeq
After some {\em unitary} interaction, the combined system is in
the state
\beq
|\Psi^1\9=\sum_\mu\alpha_\mu|\psi_{\rq \mu}^1\9\otimes|\psi_{\rc
\mu}^1\9,
\eeq
where $\sum_\mu |\alpha_\mu|^2=1$ and classical Liouville
densities $f_\mu\equiv|\psi_{\rc \mu}^1(x,k)|^2$ are
distinguishable, $\6\psi_{\rc\mu}^1|\psi_{\rc \nu}^1\9=0$,
$\mu\neq\nu$. (Several projectors may be lumped together to form a
single $\Pi$). Then if the outcome $\mu$ is found then the
wavefunction of the system ``collapses" to $\psi_{\rc
\mu}^1$. A realization of this
scheme depends on the possibility to produce the unitary $U$ that
is responsible for the interaction of classical and quantum
subsystems.

%\section{No-go theorem }

Koopmanian formalism   automatically takes care of the two basic
requirements for mixed dynamics. First, in the absence of
interaction the systems split into classical and quantum parts
that evolve according to the rules of the corresponding mechanics.
Second, expectation values of positive quantum operators and
classical probability densities are positive. This allows a simple
derivation of the no-go result (\cite{casa, pt} and references
therein) with the minimal number of assumptions.

 We assume the
following:

 $\bullet$ Quantum sector is described by
usual quantum mechanics.

$\bullet$ Classical sector is described by  Koopmanian dynamics,
with $x$ and $k$ promoted to commuting multiplication operators.

$\bullet$ A dynamics of the combined system is described by a
unitary evolution on the joint Hilbert space
$\cH=\cH_\rq\otimes\cH_\rc$ that is given by some unitary operator
$U$. The generator of the interaction part is $K_i$. This is our
definition of the mixed dynamics.

It follows from these definitions that classical and quantum
operators commute, in particular
\beq
[q,x]=[p,k]=[p,p_x]=\ldots=0. \label{defeq}
\eeq
This is just a standard result for operators that act on different
Hilbert spaces. The existence of the unitary dynamics on the
combined Hilbert spaces ensures that these relations hold all the
time.

First, we assume that the interaction term $K_i$ contains only
observables, i. e., $x$, $k$, $q$ and $p$. Then the commutativity
of the classical observables with addition of Eq.(\ref{defeq})
isolates the classical degrees of freedom from the quantum ones,
\beq
[x,K_i]=[k,K_i]=0.
\eeq
This result does not forbid a more complicated interaction term
that involves also operators $p_x$ and $p_k$. Indeed, it is
possible to construct such terms
\cite{pt}. However, to ensure that a quantum sector influences classical
observables, an interaction  $K_i(p_x,p_k,x,k,q,p)$ should have
terms of the form $qp_x$, $pp_k$ and/or similar ones. Thus the
equations for $q$ and/or $p$ will acquire terms with classical
unobservable operators, thus becoming different from both
classical and quantum equations of motion. If  formal
correspondence with either purely classical or purely quantum
equations of motion is a benchmark that is used to judge the
validity of mixed dynamics, then it fails the test.

Correspondence with the classical or quantum equations is more
than just a formal requirement. If it is violated, there will be
an inconsistency with a classical limit, as we now explain. When
the analysis is not restricted to the commuting variables,
classical behavior means that
 expectation values of operators $\hat{q}(t)$ and
$\hat{p}(t)$ tend to the expectation values of classical dynamical
variables $q(t)$ and $p(t)$ that are calculated from an
appropriate Liouville density.

More precisely \cite{bu}, POVMs that describe a joint measurement
of position and momentum result in probability distributions
\beq
f(p,q,t)dpdq=\tr[dE(p,q)\rho(t)],\label{co1}
\eeq
that go to the classical Liouville probability density $
f_\rc(p,q,t)$ that evolves as in Eq.~(\ref{Leq}). Here a POVM for
the joint position-momentum measurement satisfies $\int
dE(p,q)=1$, and $\rho$ is system's density matrix. Moreover, we
expect that
\beq
\tr(\hat{O}(t)\rho)=\6O(t)\9=\int Of(p,q,t)dqdp, \label{co2}
\eeq
where $O=q,p$, while the illusion of ``sharpness" and ``infinite
precision" of classical theory are achieved by
\beq
\Delta q/\6q\9\ll 1,\qquad \Delta p/\6p\9\ll 1\qquad \Delta q\Delta p\gg\hbar
\eeq
holding for all times. All these statements should be qualified:
the limits are expected to hold for the ``reasonable" systems and
``appropriate" POVMs. Analysis of quantum-to-classical transitions
is  a subject of current research. However, for simple systems
like two harmonic oscillators, with a bilinear coupling $cqx$,
where $c$ is a constant, the correspondence is easily established.
If we treat both of them classically, with a Hamiltonian
\beq
H=\half(q^2+p^2+x^2+k^2)+cqx \label{osc}
\eeq
we obtain equations of motion
\beq
\dot{q}=p,\qquad\qquad \dot{p}=-q-cx,
\eeq
\beq
\dot{x}=k,\qquad\qquad \dot{k}=-x-cq. \label{xy}
\eeq
 Exactly the
same equations of motion appear in the Heisenberg picture for
quantum mechanics.

 Now if for some of the variables the Koopmanian equation of motion
 $\dot{O}=\ldots$
 differs from the above, the limit of Eqs.~(\ref{co1}) and
 (\ref{co2}) will break down \cite{bal}, since
 \beq
 O(t)=O(0)+t\dot{O}(0)+\ldots.
 \eeq
 and the expectation value  $\6O(t)\9$, $t>0$ will be different from
 its classical counterpart even if they agree at $t=0$. Thus we
 see that a formal identity of the equations is necessary for the
 classical limit to hold. However, $K_i$ that is built only from
 observables cannot produce the interaction terms, while a more
 general  term cannot mimic the classical (or
 quantum) equations for observables. It will add additional terms into the
 equation for unobservable operators $p_x$ and $p_k$, which in turn will change the equations for observables
 \cite{pt}.
 For some systems, like the one given by
 Eq.~(\ref{osc}), an introduction of the the mixed dynamics leads to even more
bizarre consequences.  If the interaction term of the Hamiltonian,
$cqx$ is replaced by the most natural Koopmanian term,  $cqp_k$,
it results in the infinite flow of energy from the classical to
the quantum oscillator \cite{pt}.

This discussion can be summed as follows. To have a non-trivial
quantum-classical dynamics it is necessary to have classical
non-observable operators in the interaction term, but their
presence leads to the violation of correspondence principle and
may result in  energy non-conservation.

As a result,  it is impossible to maintain that apparatus is
described by classical mechanics through the interaction. The
``dual" approaches, where the apparatus is treated either
quantum-mechanically or classically \cite{hp,diosi} are necessary.

Since the classical gravity (pure or with matter) can be
formulated in the Hamiltonian form \cite{gravcan}, the above
result applies to it as well. Impossibility of the universal mixed
dynamics means that to get a fully consistent description of the
gravitation phenomena, there must be a quantized theory of
gravity. This analysis shows that while in each particular case
there is a specific reason for the inconsistencies or a necessity
of fine tuning, no approach that mixes a quantized matter (or
fluctuating cosmological constant) and a classical gravity is
fully consistent. Moreover,
 this adds another question
\cite{helf} to the validity of the black hole radiation and
their eventual evaporation. The radiation is derived as an effect
of a quantum field theory on a fixed curved background. The black
hole evaporation is concluded from the analysis of the
backreaction of the quantum fields on the classical metric
\cite{blackev}, which is inconsistent.  However, a conclusive
resolution of this problem will be possible only after a full
theory of quantum gravity emerges.

\bigskip

\acknowledgments

Useful discussions with Daniel Gottesman, Netanel Lindner, John
Moffat, Asher Peres, Lee Smolin, Rob Spekkens,  and Rafael Sorkin
are gratefully acknowledged.

\end{document}